# Electrochemical DNA Hairpin Sensors for Differentiating Small Molecule Intercalation from Minor Groove Binding

Kiara Thompson,[1,γ] Kian Barnes[2,γ], Saron Yoseph[3], Vincent Clark[4], J.D. Mahlum[4], Ananya Majumdar[5], Philip S Lukeman[6,] Netzahualcóyotl Arroyo-Currás[1,2,3,4,*]

[1] Program in Molecular Biophysics, Zanvyl Krieger School of Arts & Sciences, Johns Hopkins University, Baltimore, MD 21218, United States

[2] Department of Chemistry, University of North Carolina at Chapel Hill, Chapel Hill, North Carolina 27599, United States

[3] Department of Physiology, Pharmacology & Therapeutics, Johns Hopkins School of Medicine, Baltimore, MD 21205, United States

[4] Chemistry-Biology Interface Program, Zanvyl Krieger School of Arts & Sciences, Johns Hopkins University, Baltimore, MD 21218, United States

[5] Biomolecular NMR Center, Johns Hopkins University, Baltimore, Maryland 21218, United States

[6] Department of Chemistry, St. John's University, Queens, NY 11372 USA

[γ] Denotes equal contribution.

* **Correspondence to:**
Netzahualcóyotl Arroyo-Currás (Netz Arroyo)
Current Address:
Caudill Labs, Room 322
University of North Carolina at Chapel Hill
131 South Road
Chapel Hill, NC 27514
narroyo@unc.edu

## ABSTRACT

Small molecule double-stranded DNA intercalators have significant potential for therapeutic applications. However, screening for and confirming a drug candidate's intercalative behavior remains labor-intensive and costly. To address this, we investigated the sequence and biophysical parameters that affect the performance of electrochemical DNA hairpin sensors for streamlined identification of structural intercalators. These sensors utilize oligonucleotide (oligo) sequences that form hairpins upon intercalator binding. The 3' end of the oligo is modified with alkylthiol linkers for gold electrode surface monolayer self-assembly, while the 5' end carries a methylene blue redox reporter. Hairpin formation enhances electron transfer between methylene blue and the gold electrode, which can be detected via voltammetry. We tested seven hairpin structures varying in stem length and sequence. Our optimal oligo, HP4, features a four-base-pair stem and responds to five DNA intercalators over a broad detection range, with $EC_{50}$ in close agreement with published affinity ($K_D$) values for these interactions. We further demonstrate HP4's ability to discriminate intercalator binding from a series of minor groove binders through significant differences in signal gain upon incubation. Altogether, our strategy establishes a platform for identifying intercalative compounds that should support the development of DNA-targeting therapeutics.

**Key Words:** nucleic acid-based electrochemical sensors, intercalation, SYBR gold, minor groove binder, voltammetry, self-assembly, drug discovery, biosensor

## INTRODUCTION

Intercalators are small molecules capable of inserting themselves between the base pairs of double-helical nucleotide-based polymers, the most common of which is double-stranded DNA (dsDNA).[1] Intercalation alters the secondary structure through elongation[2-4] and usually enhances the thermal stability of dsDNA[5]. These structural and energetic alterations, in turn, impact dsDNA-protein interactions. While such perturbations can significantly disrupt cellular function and contribute to pathological processes, they also offer opportunities for therapeutic intervention. For instance, FDA-approved intercalating anthracyclines such as doxorubicin, epirubicin, and valrubicin are used in treating solid tumors and hematologic cancers,[6] while proflavine has served as an antiseptic.[7, 8] Although general structural principles help predict a compound's intercalation potential during drug development, confirming intercalation usually is an intensive process requiring significant cost, instrumentation, and labor. The complexity in identifying intercalation as a distinct mode of dsDNA binding partly arises from the need to differentiate it from other binding modes (e.g, minor groove binding, electrostatic interactions, etc.).

Benchmark methods for identifying intercalation as a DNA binding mode rely on direct ensemble structural confirmation techniques like X-ray crystallography[9] or nuclear magnetic resonance (NMR)[10]. Additionally, single-molecule methods, such as optical tweezers[4, 11] and force microscopy[1, 12], can directly reveal intercalation. However, these techniques are technically difficult and time-consuming, requiring, for instance, the successful crystallization of an intercalator-dsDNA complex, or they depend on access to high-field NMR or high-resolution microscopy equipment. Consequently, they are not well-suited for rapid screening of compound libraries. In contrast, indirect methods are more amenable to faster compound screening during drug development. These include viscosity measurements, which reflect changes in the structural rigidity of dsDNA due to intercalation,[13] thermal denaturation studies that indicate intercalation through an increased dsDNA melting temperature,[14] and optical spectroscopy methods, such as UV absorption, fluorescence emission, and circular dichroism, which report spectroscopic changes induced by binding.[15-17] However, these indirect techniques can also respond to other small molecule-dsDNA binding modes[1], necessitating follow-up confirmation using a direct method for greater confidence in binding-mode assignment.

We describe here an alternative technique utilizing nucleic acid-based electrochemical sensors (NBEs) that identify intercalators more rapidly than conventional direct methods, in a manner adaptable for drug screening. While previous studies have reported the use of NBEs for intercalator sensing,[18-21] this research builds upon that foundation to address two unanswered questions: (**i**) What are the minimal design requirements for an oligonucleotide (oligo) hairpin that can form NBEs for selective intercalator detection? (**ii**) Based on this minimalistic design, what factors influence the maximum achievable signal gain from such NBEs?

To answer these questions, we first synthesized a series of designed DNA hairpins and compared their DSC and $^{1}$H NMR DNA melting behavior and electrode-bound electrochemical behavior. This approach led to a hairpin design that maximized sensor signal output over a broad intercalator concentration range. The hairpin also showed strong discrimination, in the form of a nearly threefold difference in maximum signal gain for intercalators relative to minor groove binders. Given these results, the platform we propose should serve as a valuable aid in the evaluation of intercalator agents during drug discovery and development studies.

## METHODS

### Chemicals and Materials

Trizma base (PN T1503-1KG), 6-Mercapto-1-hexanol (MCH, PN 725226-1G), Hoechst 33258 solution (Hoechst, PN 94403-1), Ethidium Bromide solution (Ethidium Bromide, PN E1510-10), Doxorubicin Hydrochloride (PHR1789-200MG) were purchased from Sigma-Aldrich (St. Louis, MO). SYBR Gold nucleic acid gel stain (SYBR gold, PN S11494), procaine hydrochloride (procaine, PN 207311000) Phosphate buffered saline (1X PBS, 11.9 mM $HPO_3^{2-}$ ; 137 mM NaCl; 2.7 mM KCl; pH = 7.4), sulfuric acid ($H_2SO_4$), sodium hydroxide (NaOH) were purchased from Fisher-Scientific (Waltham, MA). Ellipticine HCl (A354526), mitoxantrone (A207951), furamidine 2HCl (A508819), and propamidine isethionate (A414514) were purchased from Ambeed (Buffalo Grove, IL). All aqueous solutions were prepared using deionized water from a Milli-Q Direct purification system, with a resistivity of 18 MΩ. Gold disk electrodes (PN 002314, diameter 1.6 mm), and coiled platinum wire counter electrodes (PN 012961) were purchased from ALS Inc. (Tokyo, Japan). Silver/silver chloride reference electrodes (PN CHI111) were purchased from CH instruments (Austin, TX). For polishing electrodes, alumina slurry (PN CF-1050) and cloth pads (PN MF-1040) were purchased from BASi (West Lafayette, IN). All DNA sequences employed in the study were purchased from Sigma-Aldrich modified on their 3’ end with methylene blue and on their 5’ end with a C6 thiol linker.

**Table 1. DNA sequences used in this study.**

| Name | Sequence |
|---|---|
| **HP1_18dT** | 5′-SH-CTT TTT TTT TTT TTT TTT TG-MB-3′ |
| **HP2_18dT** | 5′-SH-CGT TTT TTT TTT TTT TTT TTC G-MB-3′ |
| **HP3_18dT** | 5′-SH-CGA TTT TTT TTT TTT TTT TTT TCG-MB-3′ |
| **HP4_5dT** | 5′-SH-CGA TTT TTT ATC G-MB-3′ |
| **HP4_18dT** | 5'-SH-CGA TTT TTT TTT TTT TTT TTT TAT CG-MB-3′ |
| **HP5_10dT** | 5′-SH-CGA TCT TTT TTT TTT GAT CG-MB-3′ |
| **HP5_18dT** | 5'-SH-CGA TCT TTT TTT TTT TTT TTT TTG ATC G-MB-3' |
| **HP6_18dT** | 5'-SH-CGA TCG TTT TTT TTT TTT TTT TTT CGA TCG-MB-3' |
| **HP7_5dT** | 5'-SH-CGA TCG TTT TTT ACG ATC G-MB-3' |
| **HP7_10dT** | 5'-SH-CGA TCG TTT TTT TTT TTA CGA TCG-MB-3' |
| **HP7_18dT** | 5'-SH-CGA TCG TTT TTT TTT TTT TTT TTT TAC GAT CG-MB-3' |
| **HP8_18dT** | 5'-SH-CGA TCG TAT TTT TTT TTT TTT TTT TTT TAC GAT CG-MB-3' |
| **HP10_18dT** | 5'-SH-CGA TCG TAT ATT TTT TTT TTT TTT TTT TTA TAC GAT CG-MB-3' |

**Determination of SYBR Gold Concentration**

SYBR Gold 10,000X concentrate was purchased from Fisher-Scientific. Concentration determination was based on the parameters defined by Kolbeck et. al.[22] The $A_{486}$ was measured for a 10X concentrate solution using an Implen Nanophotometer NP80. Stock concentration was determined using the Beer-Lambert law with a molar extinction coefficient of 57,000 $M^{-1}$ $cm^{-1}$.

**NBE Sensor Preparation**

To prepare DNA solutions, 1 µL of ssDNA (100 µM in water) was added to 2 µL of a TCEP solution (5 mM in water) for 1 h to reduce disulfide bonds. Following the hour, DNA was diluted in PBS to a final concentration of 500 nM (or other as indicated in the manuscript) determined via an Implen Nanophotometer NP80. While DNA is reducing, electrodes were polished for ~3 min using a cloth pad with alumina slurry. After polishing, the electrodes were cycled via cyclic voltammetry in 0.5 M NaOH for 400 scans then electrochemically cycled in 0.5 M $H_2SO_4$ for 400 scans following previously published protocols.[23, 24] Electrodes were then immediately placed into 500 nM solutions of reduced DNA for 60 min at 4 °C unless deposition concentration was a stated variable. Following DNA incubation, electrodes were placed into freshly made solution of 1 mM MCH overnight.

**Electrochemical Measurements**

CH Instruments Electrochemical Analyzer (CHI 1040C, Austin, TX) multichannel potentiostat and associated software were used for all electrochemical measurements. A three-electrode cell configuration consisting of a working electrode, a coiled platinum wire counter electrode, and a Ag/AgCl reference electrode were used. Square wave voltammetry measurements were performed at various frequencies indicated in frequency maps, or at 70 and 700 Hz for dose-response curves, from -0.1 V to -0.55 V (vs Ag|AgCl), amplitude of 50 mV and a quiet time of 2 s.

**Frequency Map Characterization of HP4 Temperature Dependence**

Sensors were prepared from solutions containing 250 nM of their respective DNA following the above protocol. Measurements were performed in 100 mM Tris HCl (pH 7.4) buffer. Electrodes were incubated in their stated temperature and concentration conditions for 20 minutes prior to measurement. Square wave voltammetry measurements were from 0.0 V to -0.4 V (vs Ag|AgCl), amplitude of 25 mV and a quiet time of 2 s. Temperature control was done using a Huber CC-K6 refrigerated heating circulation bath attached to a jacketed glass electrochemical cell (Figure S1).

**Titrations of DNA Intercalators and Minor Groove Binders.** Sensors were prepared as previously stated using a deposition concentration of 200 nM HP4_18dT (or other as noted in the text). Measurements were performed in 10mL of 100 mM Tris HCl buffer (pH 7.4) containing 1% DMSO (v/v). DMSO was necessary to dissolve the highly hydrophobic intercalator compounds. Binding agent stocks were prepared in identical solution. Square wave voltammetry measurements were taken at 60 second intervals. Square wave voltammetry measurements were performed at 70 and 700 Hz, from -0.1 V to -0.55 V (vs Ag|AgCl), amplitude of 50 mV and a quiet time of 2 s. Circulator temperature was 20 °C. Sensor gain is reported as the percent change of the kinetic differential measurement[25] at the given target concentration relative to the absence of target.

**Data Analysis.** To process the files generated during voltametric interrogation, we used a Python-based custom script previously reported by our group (SACMES).[26]

**Differential Scanning Calorimetry Measurements.** TA Instruments (New Castle, DE) Nano DSC and associated software was used for all differential scanning calorimetry measurements. Cell was loaded with 50 μM DNA in 1x PBS, pH 7.4 buffer, and ran with a 1 C°/min step size.

**NMR.** NMR measurements were conducted in 1X PBS, 5% $D_2O$, at 800 μM concentration of oligo. $^1$H NMR spectra were acquired using the standard Bruker library pulse sequence zgesfpgp employing excitation sculpting[27] with an in-house water flipback modification to enhance sensitivity by minimizing exchange broadening of imino proton signals (pulse sequence available upon request). Experimental parameters used for all data sets were as follows: 600 MHz ($^1$H) spectrometer frequency; 64 scans/free induction decay (FID, in addition to 8 steady state or dummy scans); 29.7 ppm spectra width; $^1$H carrier on $H_2O$ resonance frequency; 7,142 complex data points/FID; 0.2 ms acquisition time/FID; inter-scan delay = 1.5 s; total time for each data set: 2 min. A 9 ms Eburp1 pulse was used for the initial $H_2O$-flip back and 1 ms rectangular pulse was used for subsequent excitation sculpting. Experimental parameters were optimized for maximum sensitivity in the imino proton region of the NMR spectrum.

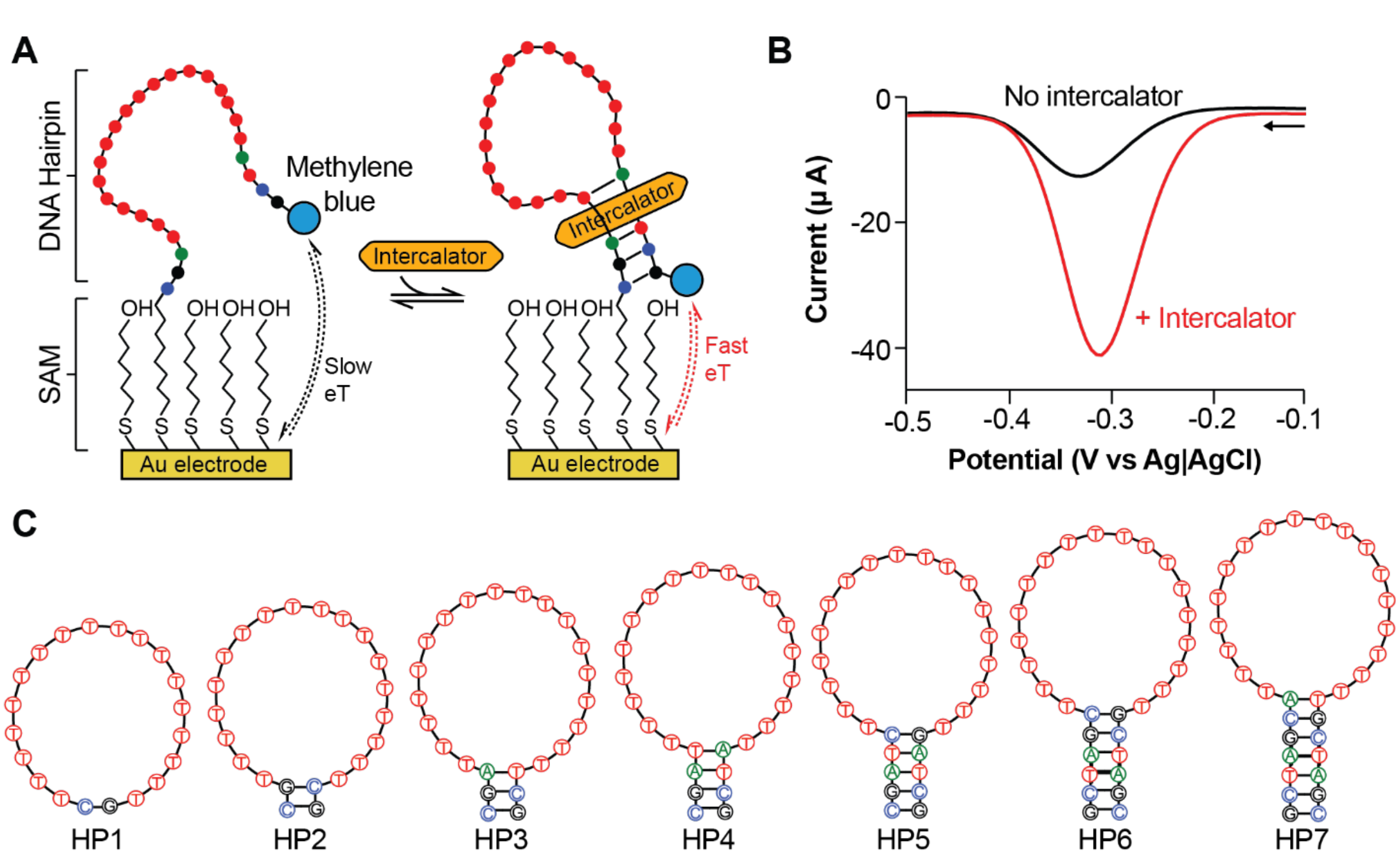

**Figure 1: Sensor design and oligonucleotide sequences tested.** (**A**) The sensors consist of gold electrodes coated with a mixed self-assembled monolayer of 6-mercapto-1-hexanol plus alkylthiol- and redox reporter-modified oligonucleotides (oligos). The oligos are designed to present complementarity sequences at their distal ends and exist in a dynamic equilibrium between folded and unfolded states on the sensor surface. Addition of an intercalator shifts the equilibrium to the folded state (right of the scheme). (**B**) Rate of electron transfer from the redox reporter to the gold electrode is an exponential function of distance. In this system, hairpin formation brings the reporter closer to the electrode surface,

thereby generating a larger current when measured via square wave voltammetry. (**C**) The hairpins considered in this study ranged from 1 to 7 base pairs in their stem. For some control measurements, we included longer stems where indicated in the text (Table 1 lists all sequences). All oligos included an 18 deoxythymidine-long loop, except for controls noted where appropriate in the manuscript.

## RESULTS AND DISCUSSION

NBEs represent a diverse family of electrochemical sensors distinguishable by variations in their biosensing interface design and structural characteristics of their affinity receptors. In this study, we developed a sensor that displays an oligo, and when exposed to an intercalator, that oligo folds into a hairpin (Figure 1A). To achieve this, oligos were synthesized with a 5′ hexylthiol linker, enabling their spontaneous self-assembly onto gold electrodes. Additionally, a methylene blue modifier was attached to the 3′ end to facilitate reversible electron transfer to these electrodes. Using square wave voltammetry, we examined the performance of various NBEs, seeking a monotonic change in voltammogram current magnitude with increasing intercalator concentrations (Figure 1B).

To identify the optimal structure for sensing, which we define as maximal signal gain and the lowest effective concentration at gain = 50% ($EC_{50}$), we evaluated a series of hairpin configurations depicted in Figure 1C. The naming convention used throughout this work designates each structure with the acronym "HP" (for hairpin), followed by a number representing the stem length in base pairs. Unless otherwise specified, all structures incorporated an 18-deoxythymidine (dT) loop to ensure hairpin formation, with deviations noted where applicable in the manuscript. We chose this design to ensure that minimal structure was imparted by the poly-dT loop, with enough degrees of freedom to fold with similar facility irrespective of stem length (per Nupack[28] predictions).

We first investigated whether methylene blue, a commonly used redox reporter in NBEs and a known intercalator[29, 30], might affect the melting temperature of the hairpin constructs we intended to use for sensor development. To investigate this, we conducted differential scanning calorimetry (DSC) measurements to determine if methylene blue, either in its free form or covalently bound to oligos, affects the melting temperature of such oligos in buffered media. For these experiments, we used an HP10 oligo (not shown in Figure 1; see Table 1 in Methods for all sequences considered in this study), with a melting temperature of approximately 65° C based on secondary structure predictions by Nupack[28] and experimentally confirmed via DSC (black trace in Figure 2A, $T_M$ = 65° ± 1º C). Adding 500 µM methylene blue to the solution (i.e., a six fold excess) increased the melting temperature by 5° ± 1º C (red trace), consistent with the expected stabilizing effect of intercalation on the stem. However, when the experiment was repeated using an oligo modified with a covalently linked methylene blue, we observed a 10° ± 1º C decrease in melting temperature (magenta trace, HP10 MB [bound]), indicating destabilization of the hairpin's stem. We have previously observed a similar effect when linking methylene blue to a vancomycin-binding aptamer[23] via a decrease in affinity by isothermal titration calorimetry (see SI file in the referenced work). Based on these findings, we conclude that the methylene blue linker used by the manufacturer (Sigma-Aldrich) in our modified oligos lowers their melting temperature, indicating a greater tendency toward the unfolded state compared to unmodified sequences. This effect is important for NBE receptor design, as sensing relies on structure switching (Figure 2A), which depends on the folding and unfolding kinetics of the oligos at a given temperature. The methylene blue linker appears to alter these kinetics by destabilizing the folded stem structure. Therefore, this destabilizing effect must be considered when designing oligos for optimal sensor performance.

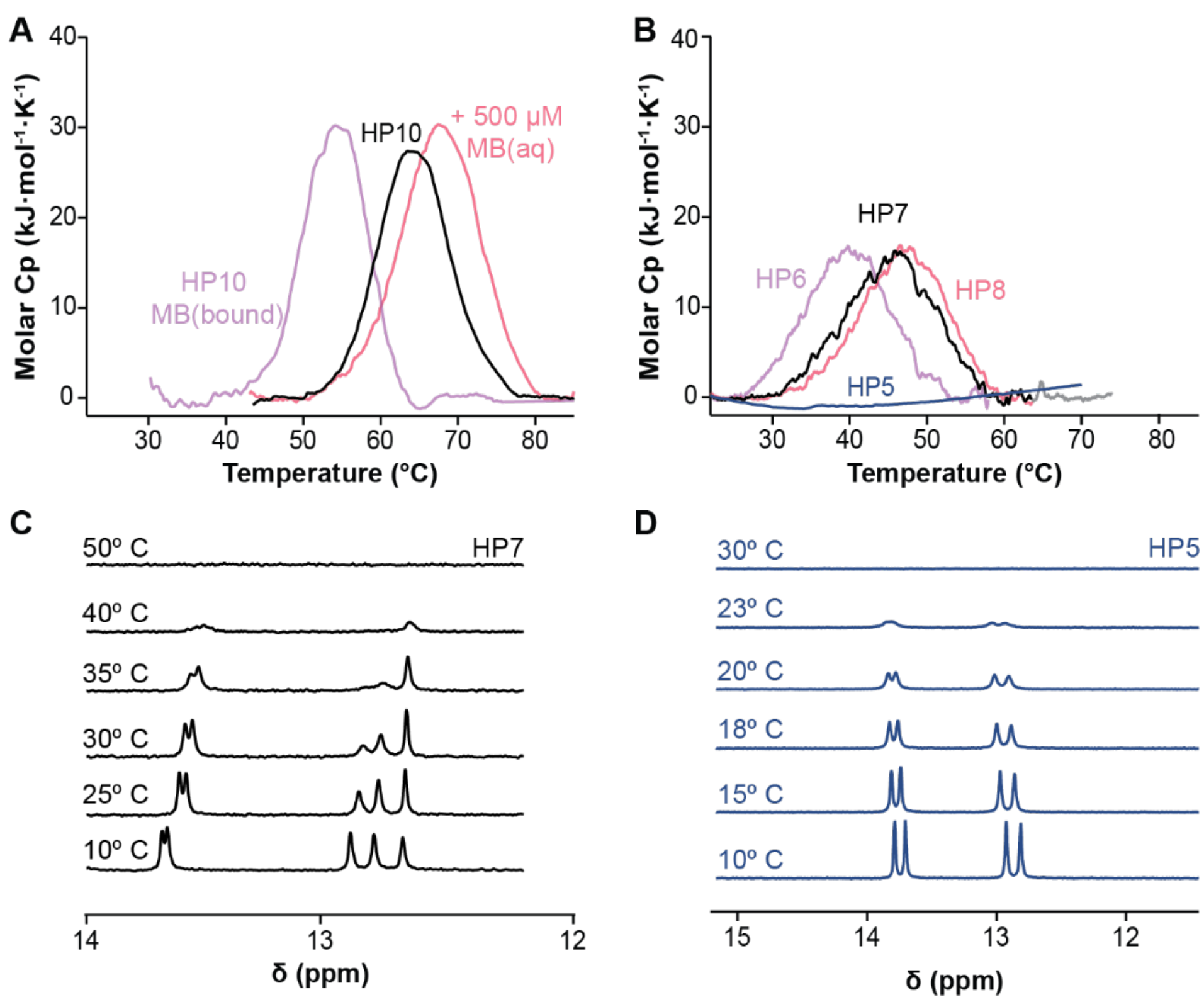

**Figure 2: Use of DSC and NMR to determine stem length stability.** (**A**) Using HP10 alone (black trace) the heat of strand unfolding was determined to be ~65° ± 1º C. When in the presence of solvated methylene blue, a known intercalator (red trace), the melting temperature increased to 70° ± 1º C. In the presence of methylene blue covalently bound to HP10 (magenta trace), melting temperature decreases to ~53° ± 1º C. (**B**) From this series, it was determined that HP5 (blue trace) was the stem length without a clear melting transition in buffered solution (i.e., no peaks are observed). (**C**) NMR spectra of HP7 across the imino region as a function of temperature. (**D**) NMR spectra of HP5 across the imino region as a function of temperature.

Next, we aimed to determine the minimum stem length with a measurable population in solution of folded state at a temperature that is convenient to assess intercalation across lead compound libraries in vitro (here chosen to be 20º C). This length is critical for designing functional structure-switching receptors, which are indispensable for the sensing mechanism of our NBEs, as structure switching modulates the position of the redox reporter relative to the electrode surface (Figure 1A). Starting with an HP8 oligo, we systematically shortened the stem by removing one base pair at a time from the base of the 18-deoxythymidine loop (structures in Figure 1C). We then monitored the impact of these truncations on the hairpins' melting temperatures via DSC (Figure 2B). Through these measurements, we identified HP6 as the shortest hairpin capable of exhibiting a measurable melting transition under our electrolyte conditions (phosphate-buffered saline, pH = 7.4). In contrast, the HP5 oligo showed no detectable transitions across the tested temperature range (22°C to 75°C), indicating that its structure was too dynamic to show a transition under DSC conditions. However, DNA binding affinity has been shown to significantly differ in solution as compared to surface-bound systems such as NBEs.[31] These behavioral differences posed the question: Could NBEs with a shorter stem than 5 bp generate a larger gain because the structural equilibrium has a preference for an unfolded state?

To support our DSC analyses with structural data, we used nuclear magnetic resonance (NMR) spectroscopy. Consistent with our DSC findings, a stable stem (such as that formed by the HP7 oligo) exhibited strong NMR signals in the imino region,[32-35] indicative of tightly paired imino protons along the stem (Figure 2C). These imino protons became increasingly solvent-accessible as evidenced by peak disappearance (a.k.a. exchange-broadening), reflecting a shift in equilibrium wherein a higher population of the imino protons are "resident" on $H_2O$ relative to the nucleic acid. Exchange broadening is manifested in NMR spectra at temperatures above the melting temperature determined by DSC. Importantly, the DNA and electrolyte concentrations differed between the two methods, leading to a measurable discrepancy between the transition temperatures (~40 °C by NMR vs. ~48 °C by DSC). Such differences are well-documented: DSC measurements often yield higher apparent transition enthalpies and temperatures than spectroscopic methods such as NMR, due in part to differences in strand concentrations, ionic strength, and the temperature dependence of single-stranded conformations.[36] Interestingly, NMR also detected imino peaks for HP5 (Figure 2D), indicating partial stem formation that was not observed by DSC. This suggests that NMR, particularly at the oligo concentrations required for structural analysis, is more sensitive in detecting low-population folded states of hairpins relative to DSC, which reports on global thermodynamic transitions.[37]

Like HP7, HP5 also displayed increased solvent accessibility for the imino peaks with rising temperatures. To further test the limits of the NMR approach, we analyzed an HP4 oligo. However, no peaks were observed in the imino region of the spectra, likely due to greater solvent accessibility which promotes exchange broadening of the proton spectra.[38, 39] The dynamic landscape of HP4 suggests that the NMR signature of those protons is too broad to be observed. Based on these measurements and on the sensor representation shown in Figure 1, we hypothesized that we need (1) the smallest measurable amount of stem formation in solution to get a good sensor, and (2) a stem length with a population in solution just below the measurable amount via NMR. Following these hypotheses, we decided to pursue the fabrication of NBEs using HP4 as the ideal sensor oligo with additional HP constructs examined as controls where noted.

To create sensors, we purchased oligos with alkylthiol and methylene blue modifications before depositing each oligo onto gold electrodes. We measured square wave frequency maps[40] at varying temperatures and concentrations of SYBR Gold (a model intercalator)[22] to characterize HP4's behavior as an intercalator sensor. We first measured the change in frequency response as a function of temperature both in the absence (Figure 3A) and presence (Figure 3B) of SYBR Gold. In the absence of the intercalator, when transitioning from 20°C to 40°C, we observed an increase in the signal-OFF frequency displayed by the quasi-reversible maxima shifting from 30 Hz to near 50 Hz. This behavior is expected as electron transfer rate is a function of temperature and faster electron transfer should be seen as a shift in the maxima position relative to the frequency axis in the maps. At lower temperatures, we see the emergence of a second peak corresponding to a faster electron transfer process at 150 Hz. We propose that at 5°C, the lower frequency state is the proportion of HP4 that lacks secondary structure and the higher frequency state is the stem-loop structured form of HP4. Thus, the states seen at 20ºC and 40ºC correspond to the melted hairpin stem-loop as a result of the elevated temperature. In the presence of a high concentration of SYBR gold, HP4 displays a single high-frequency state (Figure 3B) that matches the position of the maxima seen for the voltammogram for HP4 taken at 5°C, which we ascribe to a stem-loop structure. Frequency maps taken at 5 degree increments from 5°C to 40°C show the gradual disappearance of the high frequency maxima from the interrogation window (Figure S2). The dynamic equilibrium between the unstructured and stem-loop state is observable as the stem-loop melts with increasing temperature.

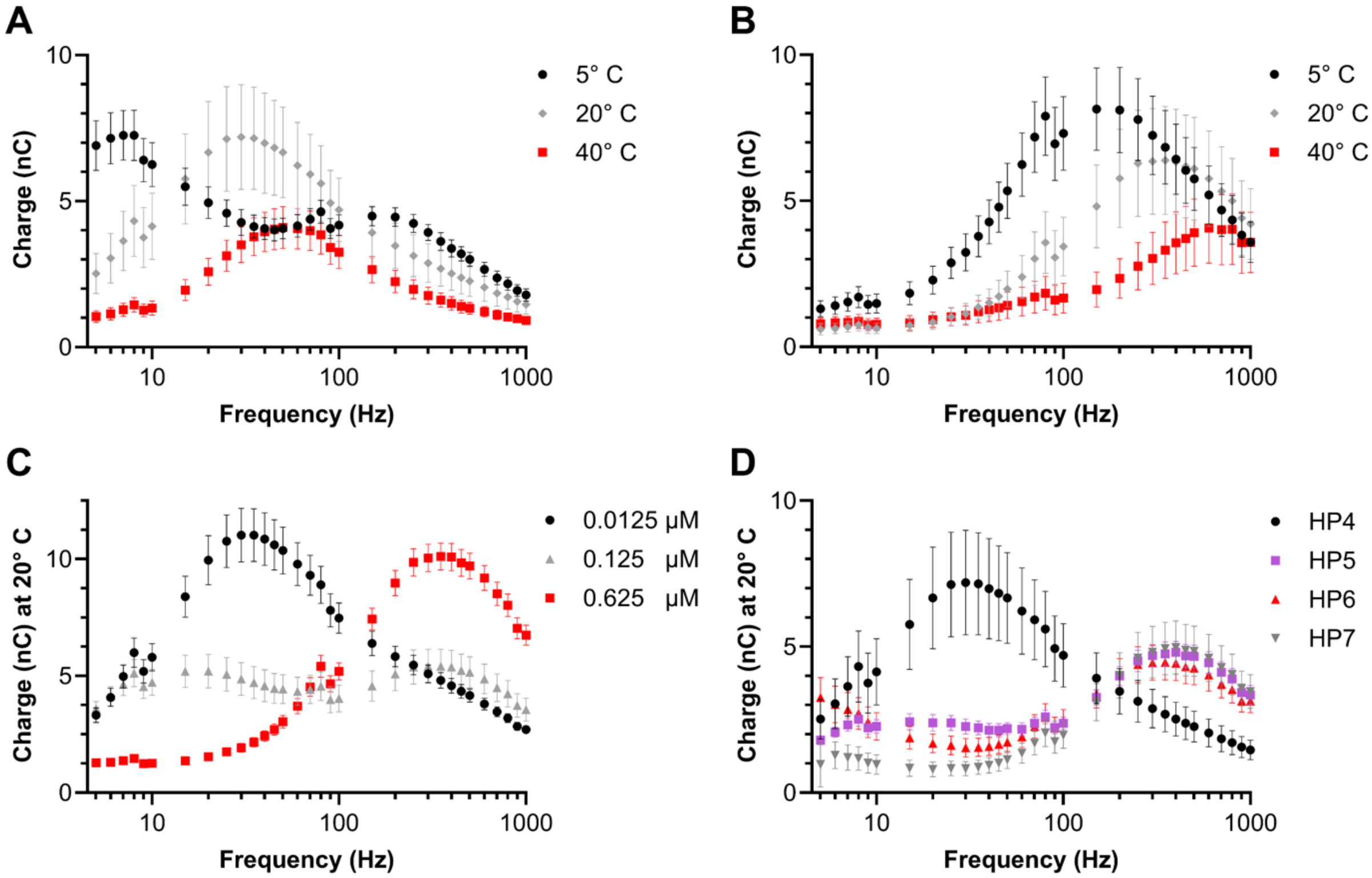

**Figure 3: Frequency map characterization of HP4 sensing behavior** (**A**) Frequency maps of HP4 in the absence of target demonstrate the charge maximum shifting to higher frequencies as a function of temperature. The unmelted HP4 stem is visualized by the high frequency peak at 5°C (black). (**B**) HP4 frequency maps in the presence of 1.25 μM SYBR Gold display a single charge maximum evidencing that the abundance of target causes the equilibrium to shift towards a stable hairpin stem. (**C**) HP4's equilibrium between its open and closed hairpin stem-loop can be perturbed with the addition of the intercalator SYBR Gold. Increasing SYBR Gold concentration at a fixed temperature of 20°C generates an increase in the high frequency maximum charge and a corresponding decrease of the low frequency maximum charge as the population shifts increasingly favoring the target-bound state. **(D)** At 20°C, HP4 predominantly lacks a stem-loop structure and displays a charge maximum at low frequency. This low frequency maximum corresponds to the hairpin's open state. Longer stem hairpins retain their stem-loop structure at 20°C and produce the majority of their signal at the corresponding high-frequency. Solid markers represent the average of 4 electrodes, error bars their standard deviation.

After establishing that modulation of the dynamic equilibrium for HP4 via temperature can be observed by measuring electron transfer rates, we also sought to visualize this modulation as the result of intercalator binding. We observed that HP4 is successfully interrogated in the presence of non-saturating concentrations of SYBR Gold, causing a change in the distribution of the two maxima at a constant temperature (Figure 3C). The relative instability of the HP4 stem-loop at 20°C makes it an effective intercalator sensor at this temperature. Intercalation significantly stabilizes the stem-loop and shifts the equilibrium distribution between open and closed hairpins. Hairpins with longer, and therefore more stable stems are presumably worse sensors at the same temperature as they are predominantly in the stem-loop high-frequency state. The elevated high frequency maxima of the longer hairpin stems - HP5, HP6, and HP7 - are clearly visible when comparing frequency maps at 20°C in the absence of target (Figure 3D). In other words, there is a lower prevalence of the unstructured state at 20°C in HP5, HP6, and HP7 represented by the lack of a low frequency maximum. The extended hairpin stems of 5 base pairs or longer are sufficiently destabilized only at higher temperatures to produce the lower frequency maximum observed with HP4 (Figure S3).

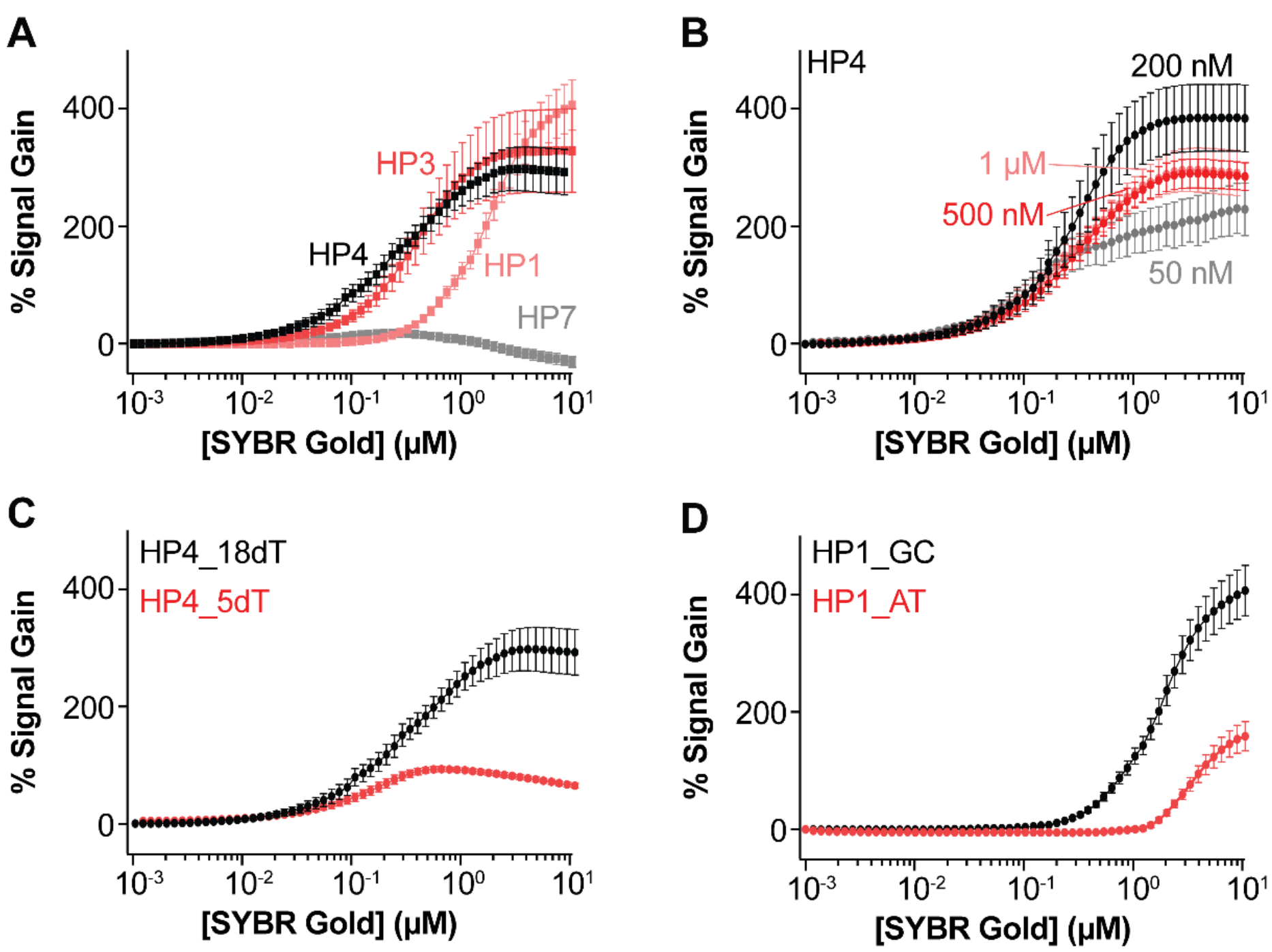

**Figure 4: Dose-response curves for SYBR Gold binding to hairpin NBEs:** (**A**) As a function of stem length, using 500 nM oligo concentration during deposition; (**B**) As a function of oligo deposition concentration using HP4; (**C**) Following 500 nM deposition, for harpins with different loop lengths; and (**D**) For HP1, deposited at 200 nM. This hairpin has a single base-pair at the base of the loop, either GC (black) or AT (red). Solid markers represent the average of 7 electrodes, errors their standard deviation.

Next, we sought to determine the best experimental conditions for NBE monitoring of intercalation leading to maximal signal gain and $EC_{50}$ values that reflect expected binding affinity of intercalators. For ease of comparison, we used frequency maps to identify common frequencies of interrogation for the various oligos and temperatures of interest. At 50°C, HP7 (the longest stem length we considered) displays a clear maximum unfolded state signal of 70 Hz, and, at 5°C, a folded state signal of 700 Hz (Figure S4). These parameters were validated for our primary oligo HP4 based on the frequency maps shown in Figure 3A-B. Using these parameters, we investigated the ability of SYBR Gold to stabilize different length hairpins and the effect of length on signal gains.

NBEs made with the dynamic HP4 oligo generated dose-response curves with gain ~ 250% at saturation (Figure 4A, black trace). Making the stem longer (i.e., going from HP4 to HP7) worsens the gain of the sensors, with NBEs functionalized with HP7 showing negative gain (grey data in Figure 4A). This is because, as shown in Figure 3D, oligos with a longer stem prefer to exist in their folded conformation at 20º C even in the absence of target; therefore, target additions do not significantly affect the relative current measured by the sensors. At this time we are not able to explain why the sensor output is negative at the higher concentrations of SYBR Gold but note this drop in signal is only a fraction of the maximum gain achieved by HP4. Alternatively, shortening the HP4 stem increases the gain to a maximum of ~400% for HP1, but with a significantly worse $EC_{50}$. Note that the NBEs response with HP4, $EC_{50}$ = 200 nM, closely matches the reported affinity constant for SYBR Gold and dsDNA in solution based on fluorescence emission measurements ($K_D$ = 273 ± 26 nM).[22]

The NBEs used to build Figure 4A were calibrated to have an average surface density of oligos = 5.0 ± 0.2 pmol/cm$^2$, as determined via integration of the current under a cyclic voltammogram of methylene blue divided by the estimation of gold electrode area per standard methods in the field.[24] Functionalizing NBEs with more densely packed oligo layers negatively affected both gain and $EC_{50}$ (Figure 4B, red data), likely due to molecular crowding effects.[41] Lower densities only decreased NBEs gain (Figure 4B, grey data). Maximal gain and lowest $EC_{50}$ were achieved by using an HP4 deposition concentration of 200 nM.

We performed additional control experiments whose results support our mechanistic hypotheses above. Challenging optimal-density HP4-functionalized sensors with volume additions of DMSO-containing buffer identical to those used during titration measurements generated negligible sensor responses (Figure S5, gain = <10%). Additionally, we evaluated the effect of loop length on the net sensor gain achievable by comparing NBEs modified with two oligos containing the same HP4 stem sequence; one with the original 18 dT loop and the other a 5 dT loop. Note that five bases in the loop should still allow the hairpin to form.[42] These measurements revealed a three-fold dampening of sensor gain for the shorter loop (Figure 4C, red data), albeit with a better $EC_{50}$ ~ 100 nM as expected. The shorter loop has a lower folding free energy and, therefore, preference for the folded form while decreasing the signal gain. On average, more hairpins are in the folded conformation at equal concentrations relative to the longer loop.

We also sought to evaluate if hairpin melting temperature could affect the maximum gain by making base pair substitutions. We hypothesized that any effect would be most prominent in NBEs containing the shortest hairpin, HP1, through the substitution of the G – C base pair for an A – T base pair. This substitution decreases the stability of the hairpin (the melting temperature) and shifts the portion of HP1 in the closed hairpin form at the given temperature. We observed two-fold reduction in gain and two-fold increase in $EC_{50}$, highlighting the importance of the interplay between hairpin stability and the maximum achievable sensor gain. However, we note that SYBR Gold intercalation to ssDNA is known to be modulated by G content, although no in the case of dsDNA[43] (as in the stem portion of our hairpin). Thus, this control may reflect the combination of G removal and stem destabilization effects.

Finally, we sought to characterize the ability of our HP4-modified NBEs to discern between modes of DNA binding. To do this, we measured the dose-response of HP4-functionalized sensors with previously characterized intercalators and minor groove binders. Ethidium bromide, a common intercalator used as a fluorescent nucleic acid stain ($K_D$ = 15 μM)[44] produces a strong sensor gain at micromolar concentrations with a maximum gain of 167 ± 14% within the titrated window (Figure 5A). Using pharmaceutically relevant intercalator targets such as doxorubicin ($K_D$ = 40 μM),[45] mitoxantrone ($K_D$ = 10 μM),[46] and ellipticine ($K_D$ = 1 μM),[47, 48] we see average maximum responses of 186 ± 8%, 237 ± 22%, and 150 ± 11% respectively (Figure 5B-D). For mitoxantrone, we interestingly observe a signal dampening at saturating concentrations. This behavior has been previously observed in other NBEs such as the vancomycin aptamer[49] and is a current topic of investigation. Relative to intercalators, the minor groove binders (Hoechst 33258,[50] procaine,[51] furamidine,[52] and propamidine[53]) generate a significantly diminished response, all below 55% signal gain, in comparison to intercalators over the same concentration window (Figure 5E-H). Using our HP4-functionalized NBEs, we demonstrate that measuring sensor gain at equivalent concentrations serves as a viable strategy for the screening of DNA binding agents. At a standard concentration of 10 μM, all intercalators studied in this work demonstrate >100% sensor gain while minor groove binders fall well below this threshold. While reported $K_D$ values for the intercalators slightly differ relative to the corresponding $EC_{50}$ values observed in Figure 5A-D, we note that those values were measured under disparate conditions that do not match our experimental conditions. Our HP4 functionalized sensors now enable lateral comparisons under identical experimental parameters, thus allowing for more convenient comparisons across libraries of intercalators.

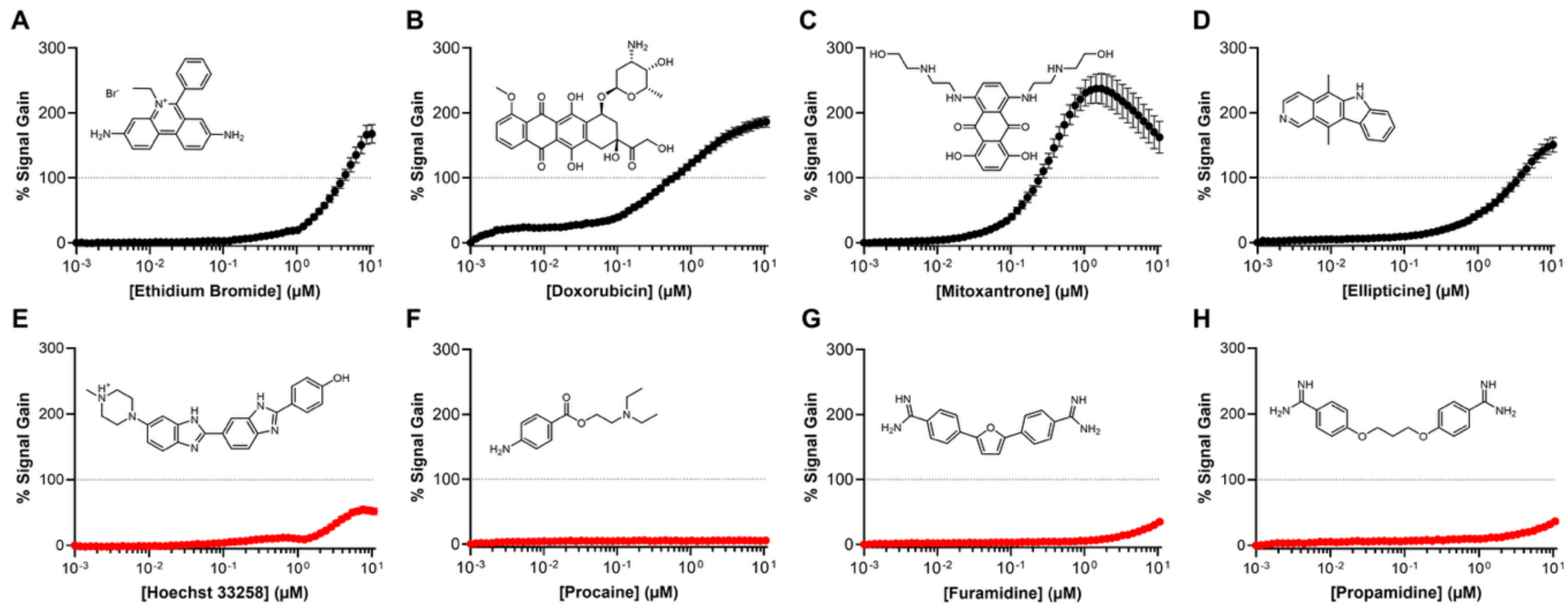

**Figure 5: Dose-response curves of HP4-functionalized sensors against intercalators and minor groove binders.** HP4 shows clear signal gains in response to titrations of intercalators **(A)** ethidium bromide, **(B)** doxorubicin, **(C)** mitoxantrone, and **(D)** ellipticine. HP4 shows diminished signal gain in response to titrating minor groove binders **(E)** Hoechst 33258, **(F)** procaine, **(G)** furamidine, and **(H)** propamidine. Solid markers represent the average of 8 electrodes, errors their standard deviation.

Overall, our results suggest HP4-functionalized NBEs may offer an excellent indirect strategy to investigate the intercalative potential of exploratory compounds during drug discovery. A single batch of modified HP4 oligos can generate thousands of NBEs, thus potentially allowing screening of compounds at a high-throughput scale via, for example, measurements across 96-well plates with an automated system.[54] Our laboratory is in the process of developing automated electrochemical analyzers to screen library compounds, which we intend to couple to the sensors discussed herein to aid drug discovery efforts.

## CONCLUSIONS

In this work, we characterize an NBE platform for the identification of DNA intercalating agents as well as to investigate the biophysical determinants of our platform. Our optimal oligo, HP4, is composed of a four base-pair hairpin structure dually modified with a 3' hexylthiol linker for gold electrode surface monolayer self-assembly and a 5' covalently linked methylene blue redox reporter. This oligo was selected based on DSC and NMR-guided analyses on how stem truncations affect its melting temperature and therefore structure-switching ability. NMR analysis of imino proton intensity as a function of temperature detected the hairpin folded state more sensitively than DSC. Our frequency maps indicate that HP4 is primarily unfolded at 20° C and visualize the changes resulting from differences in sequence architecture, temperature, and intercalator concentration. The relationship between the signal gain of our HP4-modified NBEs and the stabilization of the hairpin stem-loop structure serves as a mechanism to distinguish intercalators from minor groove binders. The large maximum signal gains generated by intercalators at micromolar concentrations are at least threefold greater than minor groove binders over the same range. Broadly, we view our HP4-based NBEs as a viable method for the early identification of DNA intercalators, and we anticipate future work to adapt this platform for high-throughput screening of DNA-intercalating therapeutic candidates.

## AUTHOR CONTRIBUTIONS

Kiara Thompson: Investigation, Data curation, Formal analysis, Writing – original draft. Performed electrochemical and NMR measurements.

Kian Barnes: Investigation, Data curation. Performed electrochemical measurements. Writing – edited draft.

Saron Yoseph: Conceptualization, Investigation. Developed initial concept and carried out experiments.

Vincent Clark: Investigation. Performed control experiments.

J.D. Mahlum: Investigation. Performed control experiments.

Ananya Majumdar: Investigation. Supported and provided resources for NMR measurements and analysis.

Philip S. Lukeman: Writing – review & editing. Contributed to data interpretation and manuscript editing.

Netzahualcóyotl Arroyo-Currás: Conceptualization, Supervision, Funding acquisition, Writing – original draft, Writing – review & editing. Directed the project, and co-wrote the manuscript.

All authors have read and agreed to the published version of the manuscript.

## CONFLICTS OF INTEREST

The authors declare no conflicts.

## DATA AVAILABILITY

Data included in the manuscript will be publicly available via deposition in the UNC Dataverse repository. A final DOI to access the data will be made available upon publication and included in the published version of this manuscript.

## ACKNOWLEDGEMENTS

A part of this study was funded by the National Institute of General Medical Sciences from the National Institutes of Health via grant R01GM140143. PSL is grateful for support from Army Research Office under Award Number W911NF-231-0283.

## SUPPLEMENTARY INFORMATION

# Electrochemical DNA Hairpin Sensors for Differentiating Small Molecule Intercalation from Minor Groove Binding

Kiara Thompson,[1,γ] Kian Barnes[2,γ], Saron Yoseph[3], Vincent Clark[4], J.D. Mahlum[4], Ananya Majumdar[5], Philip S Lukeman[6], Netzahualcóyotl Arroyo-Currás[1,2,3,4,*]

[1] Program in Molecular Biophysics, Zanvyl Krieger School of Arts & Sciences, Johns Hopkins University, Baltimore, MD 21218, United States

[2] Department of Chemistry, University of North Carolina at Chapel Hill, Chapel Hill, North Carolina 27599, United States

[3] Department of Physiology, Pharmacology & Therapeutics, Johns Hopkins School of Medicine, Baltimore, MD 21205, United States

[4] Chemistry-Biology Interface Program, Zanvyl Krieger School of Arts & Sciences, Johns Hopkins University, Baltimore, MD 21218, United States

[5] Biomolecular NMR Center, Johns Hopkins University, Baltimore, Maryland 21218, United States

[6] Department of Chemistry, St. John's University, Queens, NY 11372 USA

[γ] Denotes equal contribution.

* **Correspondence to:**
Netzahualcóyotl Arroyo-Currás (Netz Arroyo)
Current Address:
Caudill Labs, Room 322
University of North Carolina at Chapel Hill
131 South Road
Chapel Hill, NC 27514
narroyo@unc.edu

## Table of Contents

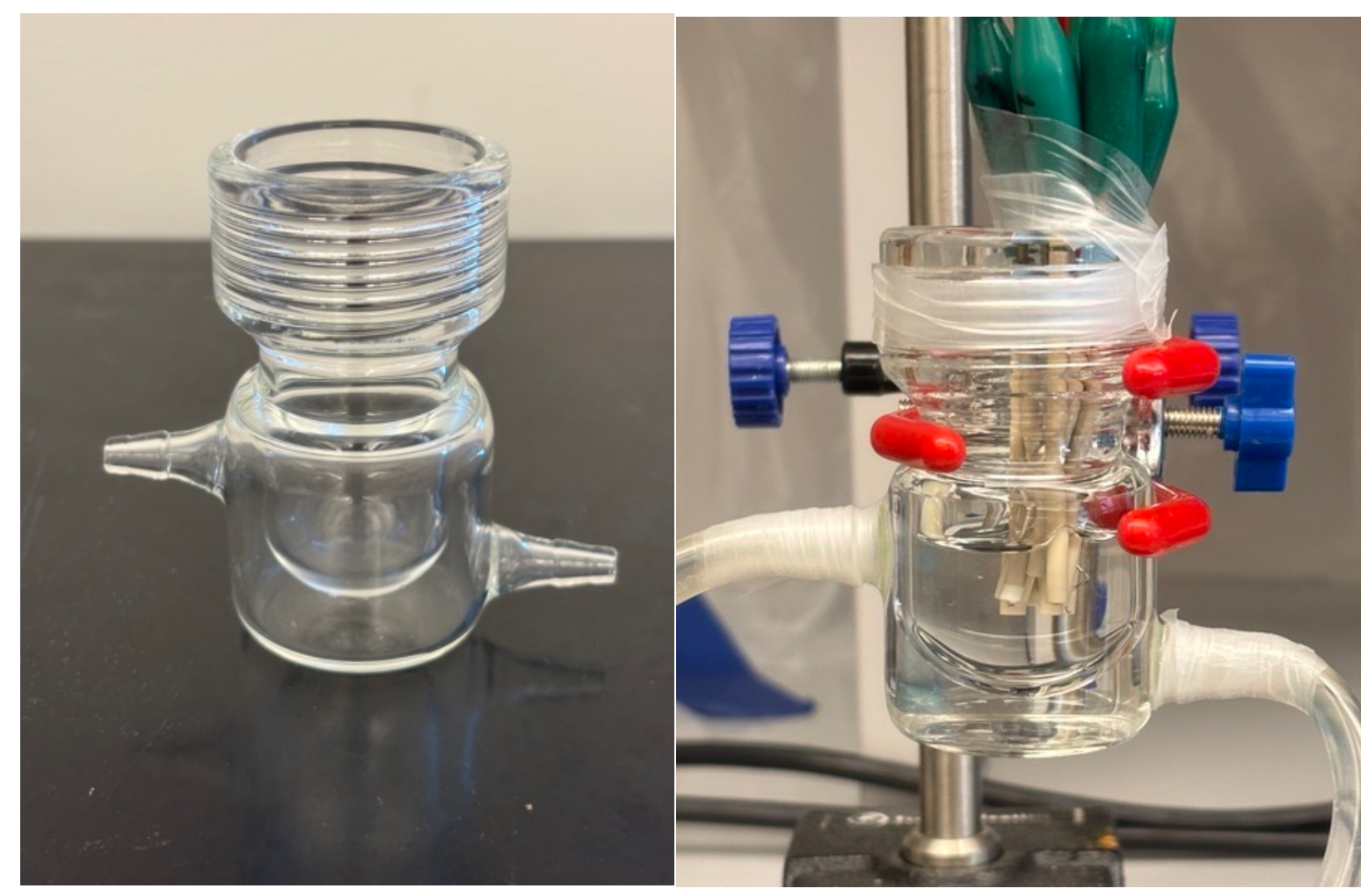

Figure S1: Jacketed glass cell used for temperature-controlled measurements

*Left:* Empty glass cell. Internal volume is 10 mL. *Right:* Full setup showing water line connections and sensors placed in blank buffered solution.

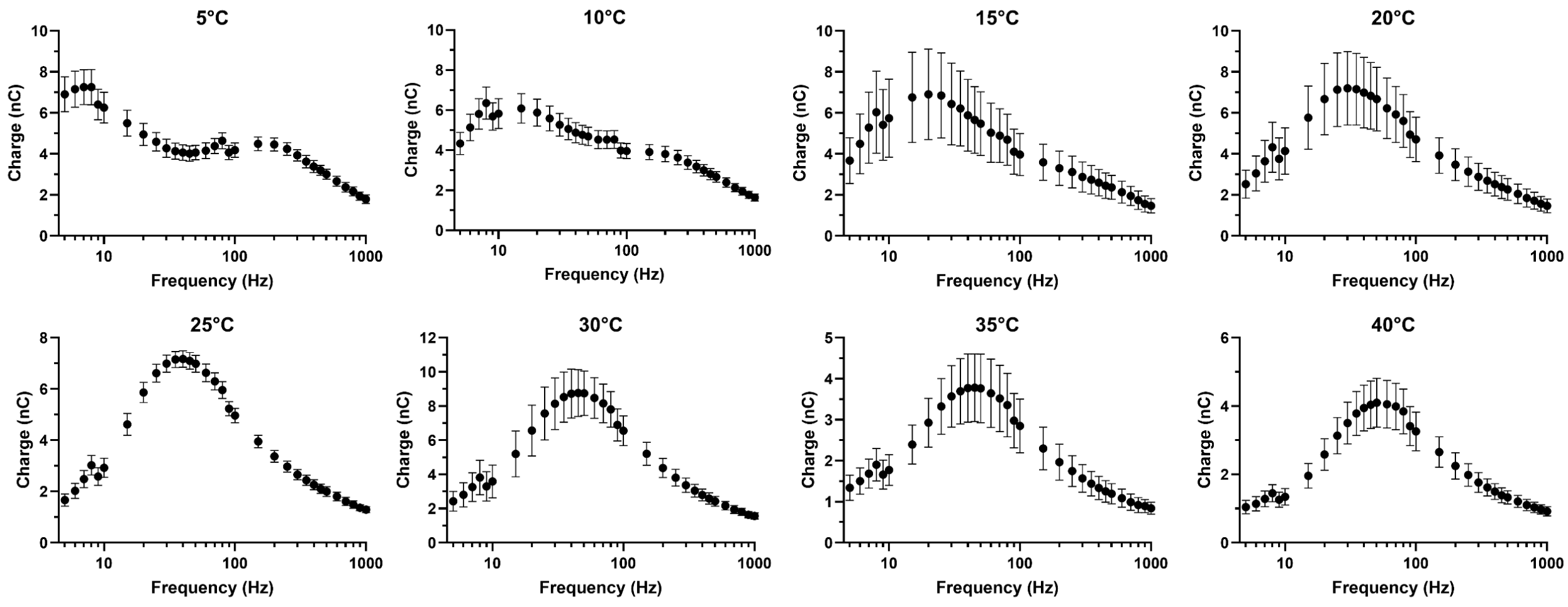

Figure S2: HP4 melting observed by temperature-resolved square wave frequency maps

Frequency maps were measured on HP4 sensors from 5°C to 40°C. Plots show the average of 4 sensors prepared in 250 nM of HP4, errors are the standard deviation. Sensors were independently prepared for each panel. Variability in error is due to small variation in DNA packing densities on the electrode surface due to day-to-day experimental variation.

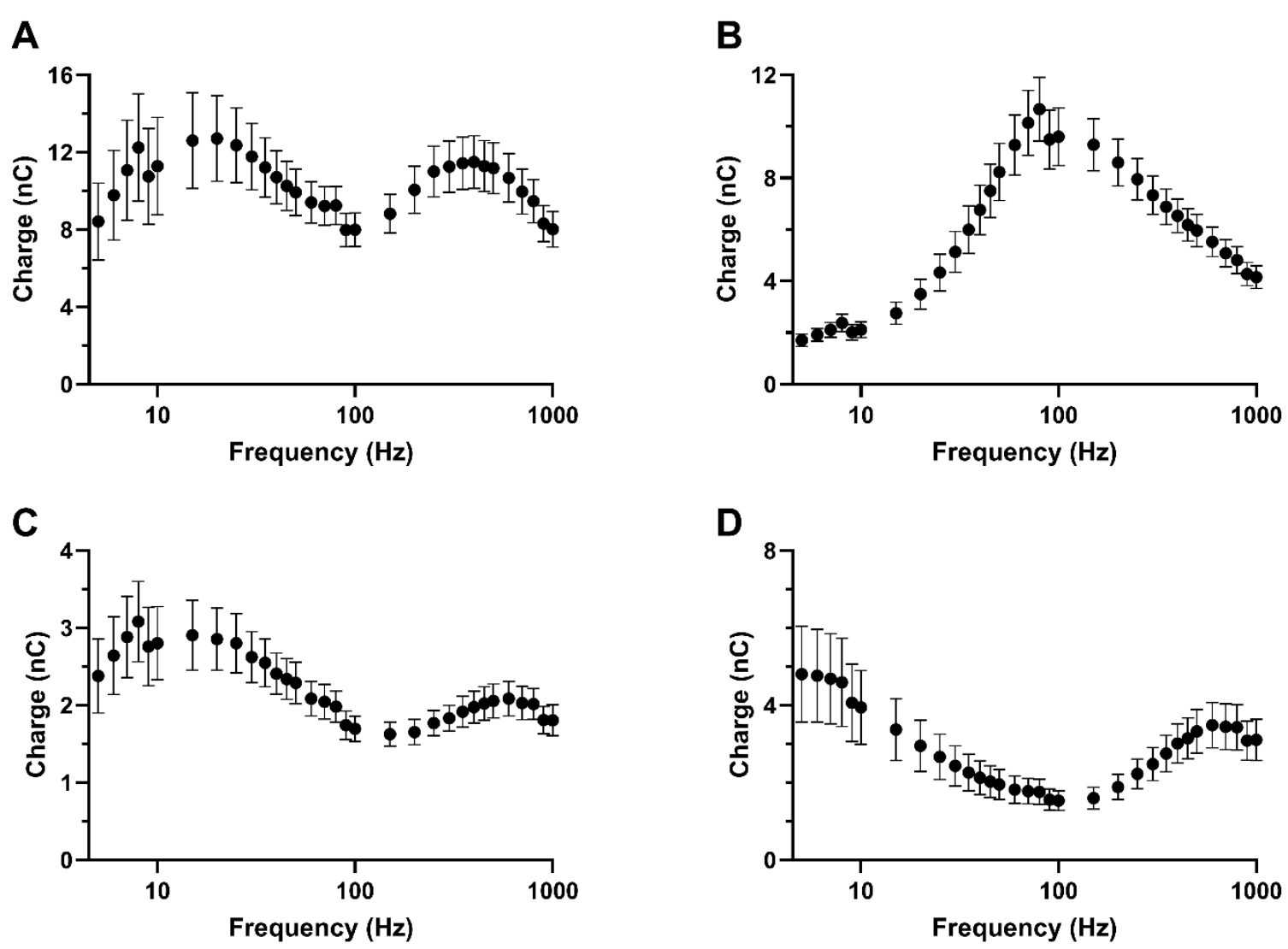

Figure S3: Frequency Map Analysis of Extended Hairpins at Elevated Temperatures

(**A**) Frequency map of HP5 performed at 25°C. (**B**) Frequency map of HP5 performed at 40°C. (**C**) Frequency map of HP6 performed at 40°C. (**D**) Frequency map of HP7 performed at 40°C. All panels show the average and standard deviation of 4 sensors independently prepared for each panel in 250 nM of their respective DNA.

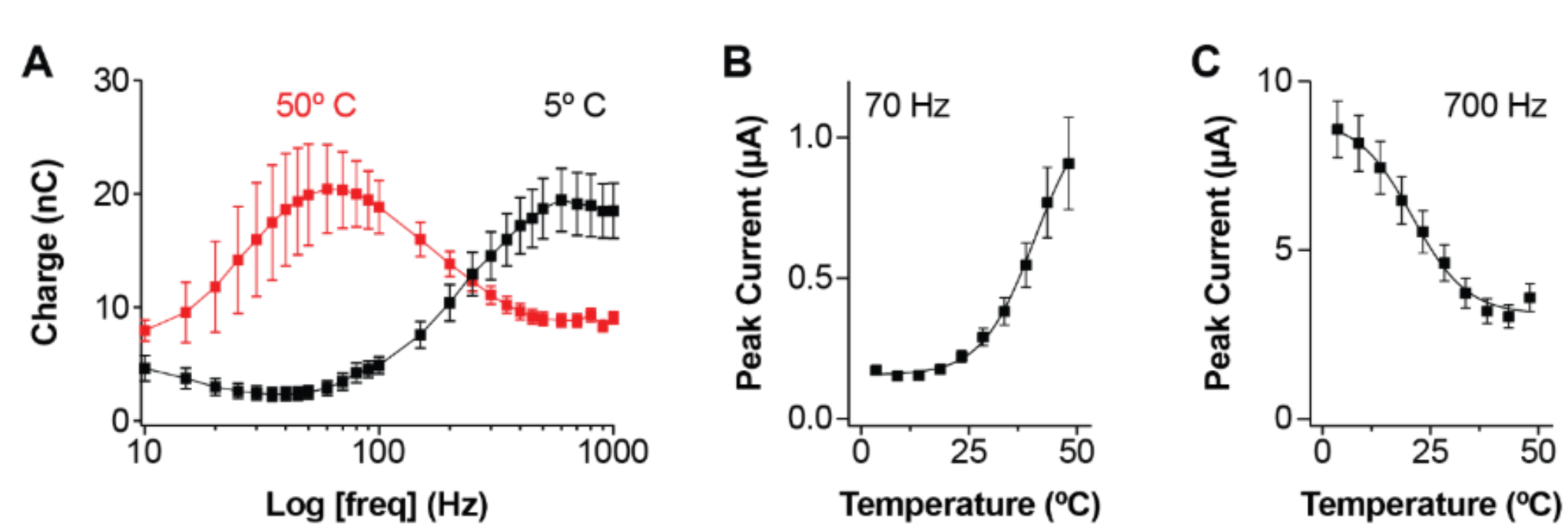

Figure S4: Identification of square wave frequencies for optimal hairpin interrogation

(**A**) To interrogate our NBEs, we first built square wave frequency maps of HP7 to determine best frequencies for interrogating slow and fast electron transferring oligo populations. (**B**) Electrochemical melting measurements of HP**7** conducted at 70 Hz reveal a small, slow electron transferring oligo fractional population that increases with melting. (**C**) Conversely, melting measurements of HP4 conducted at 700 Hz reveal an opposite trend, with a fast electron transferring population that decreases with melting. Square wave amplitude of 50 mV and step size of 1 mV for all measurements. Solid squares represent the mean of 8 independent NBEs, errors represent their standard deviation. Solid lines are point connectors**.**

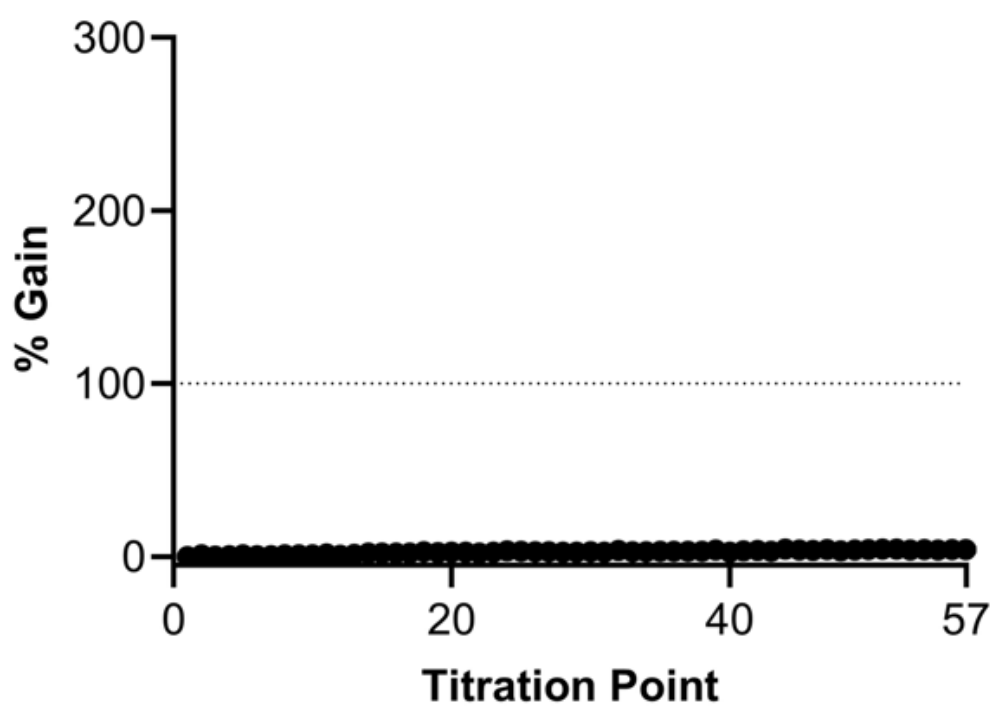

Figure S5: Control DMSO titration

Additions of DMSO in buffered solution in the same volumes and number as those employed for intercalator titrations resulted in negligible signal change for HP4-functionalized sensors.

## Table of Contents Figure

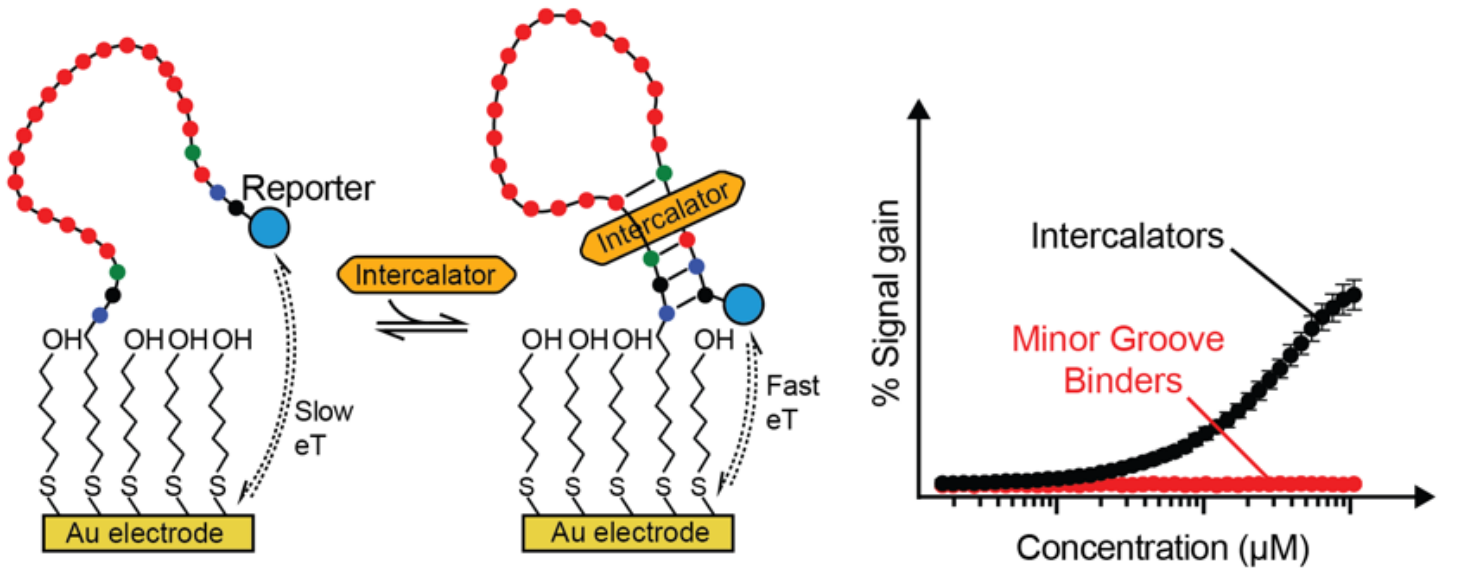